# A Changing Dichotomy: The Conception of the "Macroscopic" and "Microscopic" Worlds in the History of Physics[*]

**Zhixin Wang**, *Yale University*[†]

"Two kinds of intellectual activity, both equally instinctive, have played a prominent part in the progress of physical science," Jean Perrin wrote at the opening of his *Les Atomes* (1913).[1] The tradition of *induction* is represented by Galileo and Carnot, who were capable of establishing physical laws through the direct observations and generalizations of experience. The inquiries guided by *intuition* include the works of Dalton and Boltzmann, which inspect the "hidden gears" of the empirical world and seek to "explain the complications of the *visible* in terms of *invisible* simplicity."[2] The temporary success of one method, as Perrin opined, cannot deny the necessity of the other. Furthermore, he envisaged the day when the two approaches would merge into one—a time when individual atoms would be as visible as microorganisms.

*Les Atomes*, viewed by many as the manifesto for the reality of atomic hypothesis, was published at a critical point in the history of modern science: since the turn of the century, the crises of classical physics and the discoveries related to particles and radiation had provoked a serious quest for the mysteries of the atomic world, but the revolutionary theory, later known as quantum mechanics, would not be established in another decade. In modern vocabulary, Perrin's dichotomy between the "inductive" and "intuitive" inquiries can be rephrased as the classification of physics research according to the *scale* of the physical objects being studied. Subfields like classical mechanics and thermodynamics are thus distinct from microphysics, which involves the study of atoms, molecules, and elementary particles. However, few physicists of the following generations would fully accept Perrin's criterion—visibility—as the most sensible standard to distinguish macroscopic phenomena from microscopic ones. In fact, given the progress of modern physics since the quantum revolution, one might find this borderline increasingly hard to draw.

---

[1] Jean Perrin, *Atoms*, translated by D. Ll. Hammick from the 4th revised edition of *Les Atomes* (New York: D. Van Nostrand, 1916), pp. v.

[2] *Ibid.*, pp. vii.

This essay traces the conceptual history of *micro-* and *macroscopicity* in the context of physical science. By focusing on three distinct episodes spanning five centuries, we show the scientific and philosophical meanings of this antonym pair, despite never being far from "the small" and "the large," have been evolving as the frontier of science advances. We analyze the intellectual and material impetus for these movements, and conclude that this conceptual history reflects the changing interaction between the natural world and humankind.

## I. 17th Century: Visibility to the Unassisted Eye

Anthony Leggett remarked in *The Problems of Physics* (1987):

> To the twentieth-century scientist and the twentieth-century layman alike, few things seem more natural or self-evident than the idea that the way to understand the properties and behaviour of a complex object is to take it apart into its constituent elements.[3]

It would not be easy for modern minds to imagine a time when the study of the basic entities of matter was not only absent in scientific practice, but as some schools of thought claimed, should even be excluded from rational thinking. In Renaissance natural philosophy, "qualities"—the Aristotelian term for the causes of properties and natural effects—is divided into the "manifest" and the "occult," depending on whether they are perceptible to human senses.[4] "Manifest" properties include color, motion, weight, etc., whereas magnetism, chemistry, and medicine are all "occult" because they could only be perceived through their effects but not their mechanisms. Aristotelianism emphasizes the completeness of sensation and therefore regards the "occult qualities," which appeared to be insensible and unexplainable, as inappropriate topics for rational philosophy. "We perceive the Actions but not the qualities

---

[3] A. J. Leggett, *The Problems of Physics* (Oxford: Oxford Univ. Press, 1987), pp. 35.
[4] Keith Hutchison, "What Happened to Occult Qualities in the Scientific Revolution?" *Isis*, 1982, 73:233–253.

whereby they are affected,"—the quote from Daniel Sennert shows the prevailing impediment to modern science in the early 17th century.[5]

Such a doctrine was challenged during the Scientific Revolution. As René Descartes asserted, mechanical explanations will be found for all "occult qualities," which are therefore within the scope of scientific reasoning.[6] Moreover, he believed all perceptible effects are essentially generated by those hidden mechanisms. Meanwhile, pioneers of experimentation demonstrated that human senses can be extended through the assistance of scientific instruments. A remarkable example is the microscope, which had a major impact on both academic research and popular culture. In the Preface to his *Micrographia* (1665)—the first treatise on optical microscopy, Robert Hooke described instruments as "artificial organs" remedying the infirmities of "the Senses," and thus rectifying the operations of "the Memory, the Judgement, and the Reason."[7] Some contemporaries accepted this tool with even greater passion. For instance, after looking through Cornelis Drebbel's microscope, the exhilarated Constantijn Huygens called the view through its lenses "a new theater of nature, another world"—a world worth meticulous observations and illustrations.[8]

Although its direct contribution to the knowledge of basic physics was not yet significant, the infant microscope marked the arrival of an age when miniature structures, previously invisible, could form an extended vision for the sake of scientific observation and experimentation. Those objects that are obscure to the unassisted eye but visible with optical equipment comprise the early modern conception of the "microscopic world."

## II. *Fin de siècle*: The Struggle of Atomic Hypothesis

Owing to the fast diffusion of the microscope as a gadget for research and leisure, by the late 17th century, its magnified image was deprived of the privilege of being a "new world" but absorbed into human's natural sensation. In the meantime, the early modern period witnessed the renaissance of the ancient *atomism*—the belief that all matter is composed of undividable fundamental entities. Notably, Descartes, Boyle, and Newton all employed the model of "corpuscles" to explain astronomical motion, chemical reactions, and light, respectively.[9] The early modern *corpuscularianism* is nevertheless attached to little experimental foundation. The reality of microscopic particles was eventually to be revealed thanks to the advances of two branches of physics in the second half of the 19th centuries: In statistical mechanics, the concepts of *atoms* and *molecules* were used to derive the thermodynamic quantities of a system from its probable microscopic configurations.[10] Meanwhile, charged particles—*electrons* and *ions*—were conceived to explain the newly discovered electromagnetic phenomena, e.g. the Hall effect, the Kerr magneto-optic effect, and the Zeeman effect.[11] However, without direct experimental evidence, the true existence of these particles beyond a convenient hypothesis remained a matter of belief. For example, Ernst Mach, a firm positivist, would question whoever spoke to him of atoms: "Have you seen one?"[12]

---

[5] *Ibid*.

[6] Catherine Wilson, "Visual Surface and Visual Symbol: The Microscope and the Occult in Early Modern Science," *Journal of the History of Ideas*, 1988, 49:85–108.

[7] Robert Hooke, *Micrographia: or Some Physiological Descriptions of Minute Bodies Made by Magnifying Glasses. With Observations and Inquiries Thereupon* (London: The Royal Society, 1665), Preface.

[8] Svetlana Alpers, *The Art of Describing: Dutch Art in the Seventeenth Century* (Chicago: Univ. Chicago Press, 1983), pp. 6–7. Constantijn Huygens is a poet, diplomat, scholar, composer, and architect living in the Dutch Golden Age. He is the father of the prominent scientist Christiaan Huygens.

Hardly could one reply with a definite "yes" before Perrin's systematic study of the Brownian motion and reliable measurements of the Avogadro number in 1907–1909.[13] Around the same decade, various charged particles were observed and measured in cathode rays, cloud chambers, and falling drops, where particles are isolated from the bulk.[14] These findings preluded the profoundest reformation of the scientific knowledge of the material world in modern history, where things are "not only queerer than we suppose, but queerer than we can suppose."[15] Then came the prominence of *reductionism* in the practice of science: a complex phenomenon should be understood from the properties and configurations of its components. What was exactly opposite to the dominant philosophy before the Scientific Revolution almost became the new doctrine in the 20th century. All these transformations were initiated by the success of atomic physics.

---

[9] Their representative publications on corpuscularianism are: René Descartes, *Traité du monde et de la lumière* (1664); Robert Boyle, *The Sceptical Chymist: or Chymico-Physical Doubts & Paradoxes* (1661); Isaac Newton, *Opticks: or, A Treatise of the Reflexions, Refractions, Inflexions and Colours of Light* (1704).

[10] R. K. Pathria and Paul D. Beale, *Statistical Mechanics, 3rd edition* (Singapore: Elsevier, 2011), pp. xxi–xxvi; Stephen G. Brush, "Foundations of Statistical Mechanics 1845–1915," *Archive for History of Exact Sciences*, 1967, 4:145–183.

[11] Jed Z. Buchwald, *From Maxwell to Macrophysics: Aspects of Electromagnetic Theory in the Last Quarter of the Nineteenth Century* (Chicago: Univ. Chicago Press, 1983).

[12] S. G. Brush, "Mach and Atomism," *Synthese*, 1968, 18:192–215.

[13] Jean Perrin, *Brownian Movement and Molecular Reality*, translated by F. Soddy from the *Annales de chimie et de physique, 8me series, September 1909* (London: Taylor & Francis, 1910).

[14] O. W. Richardson, *The Electron Theory of Matter* (Cambridge: Cambridge Univ. Press, 1914).

[15] J. B. S. Haldane, *Possible Worlds and Other Essays* (London: Chatto & Windus, 1927), pp. 286.

Considering these backgrounds, especially the struggle to validate atomic reality among *fin-de-siècle* physicists and chemists, Perrin's criterion for differentiating between macro- and microphysics seems fairly reasonable. But unlike its meaning at the time of Huygens and Hooke, the "visibility" of scientific atomists does not refer to the resolution of the eye, but the very limit of experimental methods. This new notion, however, was soon to be modified when the foundation of physical science was rewritten by quantum mechanics.

### III.  Mid-20th Century Afterwards: Quantumness and More

Since the turn of the 20th century, physicists have been questing for the "most fundamental" building blocks of nature. Particles more elementary than atoms—quarks, leptons, gauge bosons, and the Higgs boson—were discovered following stringent standards.[16] Moreover, Perrin's prospect of seeing atoms under the microscope has indeed been realized.[17] Invisibility is thus no longer the distinctive feature of the atomic world. "Microscopic" objects are thereafter not characterized primarily by their spatial scale, but through the physical laws they obey: being "microscopic" means their behavior has to be described using the formalism of quantum mechanics.

Different as it may sound from its predecessor, in practice this new criterion requires very few objects to be reclassified. For a long time, quantum mechanics has been considered the strange theory for atoms, small molecules, and subatomic particles, and therefore, "microscopic," "atomic," and "quantum mechanical" more or less denote the same regime. However, this *de facto* correspondence has no solid basis: nothing in the current version of quantum theory forbids it from describing larger objects or more complicated systems. This already came into the mind of Erwin Schrödinger as early as 1935, when he raised the "cat paradox" trying to present the absurdity of applying the quantum formalism to macroscopic things and living beings.[18] This paradox has nevertheless sparked the interest among contemporary physicists to study the quantum effects of objects much larger than atoms and molecules. A few lines of work have been remarkable: First, several classes of condensed matter, e.g. superconductors, superfluids, and Bose–Einstein condensates, manifest the *macroscopic accumulation of microscopic quantum effects*.[19] Second, *quantum transport* phenomena, such as the quantized conductance and the quantum Hall effect, emerge in high-quality solid-state samples where the electron coherence length is comparable to the sample size.[20] Third, certain types of solid-state device, once sufficiently isolated from the environment, possess a quantized energy-level structure and can be prepared to a state containing no more than a few excitation quanta, and even to a coherent superposition of different low-excitation states.[21] Such engineered quantum systems are given the name "artificial atoms." They are the basic components of the *quantum machines*, whose macroscopic degrees of freedom for control and measurement are intrinsically quantum mechanical.[22] Those systems containing a large number of particles and thus "macroscopic" in spatial scale but exhibiting quantum coherence lie in the scope of *mesoscopic physics*, in which the "macro–micro" borderline is separated from the "classical–quantum" boundary.[23] In these cases, the breakdown of the one-to-one correspondence between "atomic" and "quantum mechanical" forces us to rethink the essence of the old two-world dichotomy.

---

[16] In particle physics, the standard for the "evidence of a particle" is $p = 0.003$ (three-sigma), and the standard for the "discovery of a particle" is $p = 0.0000003$ (five-sigma).

[17] For instance, individual atoms can be imaged by the scanning tunneling microscope on a conducting surface, or by the quantum gas microscope in an optical lattice.

Why does there have to be a dichotomy? In fact, more than a century ago, Perrin already conceived the merge of the two methods of inquiry. What we have been searching for is a universal set of basic principles, a consistent way of perception, that is applicable to all physical systems regardless of their spatial size or particle number. We are halfway toward this goal: On the one hand, the grand success of quantum mechanics grants us the confidence that the theory may also work at a larger scale. On the other hand, no one has ever found a "Schrödinger's cat," namely, an everyday object in a quantum superposition of classical states that are macroscopically distinct.[24] Further explorations in mesoscopic physics might provide more insightful solutions to this dilemma.

---

[18] Erwin Schrödinger, "Die gegenwärtige Situation in der Quantenmechanik," *Naturwissenschaften*, 1935, 23:807–812.

[19] James F. Annett, *Superconductivity, Superfluids, and Condensates* (Oxford: Oxford Univ. Press, 2004).

[20] Supriyo Datta, *Electronic Transport in Mesoscopic Systems* (Cambridge: Cambridge Univ. Press, 1995).

[21] D. Estève, J.-M. Raimond, and J. Dalibard, eds., *Quantum Entanglement and Information Processing*, Lecture Notes of the Les Houches Summer School, Session LXXIX (Amsterdam: Elsevier, 2004).

[22] S. M. Girvin, "Circuit QED: superconducting qubits coupled to microwave photons," in *Quantum Machines: Measurement and Control of Engineered Quantum Systems, Lecture Notes of the Les Houches Summer School, Session XCVI*, eds. M. Devoret, B. Huard, R. Schoelkopf, and L. F. Cugliandolo (Oxford: Oxford Univ. Press, 2014), pp. 113–255.

[23] Michel Devoret, "De l'atome aux machines quantiques," in *Leçons inaugurales du Collège de France* (Paris: Collège de France/Fayard, 2008).

**Conclusion**

After sketching a few key episodes in the history of physics, we find the conception of the "macroscopic" and "microscopic" worlds is by no means constant over time. Instead, it has been actively reflecting our limit of experimentation, perception, and comprehension regarding the composition and behavior of matter. Besides the development in theory and philosophy, the evolution of instrumentation plays an essential role in this odyssey. Interestingly, *the reality of atomic hypothesis was verified through the study of the Brownian motion of granules under the microscope, and the operation of a quantum machine is based on the precise control and measurement of natural or artificial atoms*. This three-stage relay is emblematic of the scientists' growing capability of observing and manipulating the subtle structures of nature throughout the ages.

In this historical review, two points deserve extra attention: the function of the measurement apparatus, which connects the "microscopic" objects to our "macroscopic" experience, and the role of the experimenter, who eventually assigns meanings to everything being observed. Meditations in this direction will recall a series of open questions, like the quantum measurement problem[25] and the influence of psychology on scientific research.[26] Some of them are recently evoked by the emergence of mesoscopic physics and quantum machines, where the line between atoms and machines is blurred, and the distance between the experimenter and the quantum world is greatly reduced. A good way to conclude this essay is to revisit Percy Bridgman's opinions back in 1958:

---

[24] A. J. Leggett, "Macroscopic Quantum Systems and the Quantum Theory of Measurement," *Supplement of the Progress of Theoretical Physics*, 1980, 69:80–100; "Quantum Mechanics at the Macroscopic Level," in *Chance and Matter, Lecture Notes of the Les Houches Summer School, Session XLVI*, eds. J. Souletie, J. Vannimenus, and R. Stora, (Amsterdam: Elsevier, 1987), pp. 395–506. For a recent review, see Florian Fröwis, Pavel Sekatski, Wolfgang Dür, Nicolas Gisin, and Nicolas Sangouard, "Macroscopic quantum states: Measures, fragility, and implementations," *Review of Modern Physics*, 2018, 90:025004. For a pedagogical introduction, see Gregg Jaeger, "What in the (quantum) world is macroscopic?" *American Journal of Physics*, 2014, 82:896–905.

[25] Eugene P. Wigner, "The Problem of Measurement," *American Journal of Physics*, 1963, 31:6–15; Maximilian Schlosshauer, "Decoherence, the measurement problem, and interpretations of quantum mechanics," *Review of Modern Physics*, 2005, 76:1267–1305.

[26] P. W. Bridgman, *The Way Things Are* (Cambridge, Massachusetts: Harvard Univ. Press, 1959), pp. 200–248.

. . . the object of knowledge and the instrument of knowledge cannot legitimately be separated, but must be taken together as one whole. . . . there is no such thing as a microscopic domain which is "revealed" to us by the microscope, but that there is rather an altered macroscopic domain, which we have found how to alter by the invention of the microscope—an obviously macroscopic instrument. . . .

Conventional quantum theory distinguishes between object of knowledge and instrument of knowledge. But knowledge itself implies a knower, . . . not only must we hold ourselves to an awareness of the microscope, but also to an awareness of ourselves using the microscope and giving its results significance. The latter is admittedly difficult, . . . but I believe we have to find how to do it before we can be satisfied that we have achieved even an approach to intellectual mastery.[27]

---

[27] P. W. Bridgman, "Remarks on Niels Bohr's Talk," *Dædalus*, 1958, 87:175–177.